\documentclass{IEEEtran}

\usepackage[export]{adjustbox}
\usepackage{graphicx}
\usepackage{graphics}
\usepackage{listings}
\usepackage{wrapfig}
\usepackage{float}
\usepackage{color}
\usepackage{times}
\usepackage{epsfig}
\usepackage{amsfonts}
\usepackage{cite}
\usepackage{subfig}
\usepackage{url}
\usepackage{mathptmx}
\usepackage{verbatim} 
\usepackage{array}
\newcolumntype{q}[1]{>{\centering\arraybackslash\hspace{0pt}}p{#1}}

\usepackage{upgreek}

\newcommand{\MYcomment}[1]{}

\newcommand{\MYnote}[1]{}

\newcommand{\MYsectionref}[1]{Section~\ref{#1}}

\newcommand{\MYfigureref}[1]{Figure~\ref{#1}}

\newcommand{\MYtableref}[1]{Table~\ref{#1}}

\newcommand{\MYcaptionfontsize}{\small }

\newcommand{\MYlabel}{\small {$\bullet$}}

\newenvironment{MYlistwide}{\begin{list}{\MYlabel}{%
\setlength{\topsep}{2pt plus 0pt minus 0pt}%
\setlength{\itemsep}{2pt plus 0pt minus 0pt}%
\setlength{\parsep}{2pt plus 0pt minus 0pt}%
\setlength{\parskip}{0pt plus 0pt minus 0pt}%
\setlength{\parindent}{0pt }
\setlength{\leftmargin}{0.14in}%
}}{\end{list}}

\newcounter{MYenumctrtwo}
\newenvironment{MYenumwide}{\begin{list}{\arabic{MYenumctrtwo}.}{%
\usecounter{MYenumctrtwo}%
\setlength{\topsep}{2pt plus 0pt minus 0pt}%
\setlength{\itemsep}{2pt plus 0pt minus 0pt}%
\setlength{\parsep}{2pt plus 0pt minus 0pt}%
\setlength{\parskip}{0pt plus 0pt minus 0pt}%
\setlength{\parindent}{0pt }
\setlength{\leftmargin}{15pt}%
}}{\end{list}}

\newcounter{MYenumctr}

\makeatletter

\begin{document}
\newcommand{\systemname}{Ubora~}

\title
{\LARGE Measuring and Managing Answer Quality\\ for Online Data-Intensive Services}
\author{
      \begin{tabular}[t]{q{2.15in}q{2.1in}q{2.0in}} 
         Jaimie Kelley, Christopher Stewart  &  Devesh
         Tiwari &Yuxiong He \& Sameh Elnikety\\
         \& Nathaniel Morris {\em The~Ohio~State~University}
         & {\em Oak Ridge National Laboratory} & {\em Microsoft Research}\\
       \end{tabular}\par\
}

\date{}

\maketitle

\thispagestyle{empty}

\begin{abstract}
\label{sect:abstract}
Online data-intensive services parallelize query
execution across distributed software components. 
Interactive response time is a priority, so online query
executions return answers without waiting for slow running
components to finish.  However, data from these slow components
could lead to better answers.  We propose Ubora, an approach to
measure the effect of slow running components on the quality
of answers.  Ubora randomly samples online 
queries and executes them twice. The first execution
elides data from slow components and provides fast online
answers; the second execution waits for all components to
complete.  
Ubora uses memoization to speed up mature executions by
replaying network messages exchanged between components.
Our systems-level implementation works for a wide range of
platforms, including Hadoop/Yarn, Apache Lucene, the EasyRec
Recommendation Engine, and the OpenEphyra question answering
system.  Ubora computes answer 
quality much faster than competing approaches that do not
use memoization. 
With Ubora, we show that answer quality can and should be
used to guide online admission control.  Our adaptive
controller processed 37\% more queries than a competing
controller guided by the rate of timeouts.
\end{abstract}

\section{Introduction}
\label{sect:intro}

{\em Online data-intensive (OLDI) services}, such as search
engines, product recommendation, sentiment analysis and Deep
QA power many popular websites and enterprise products.
Like traditional Internet services, OLDI services must
answer queries quickly.  For example, Microsoft Bing's
revenue would decrease by  \$316M if it
answered search queries 500ms slower~\cite{forrest-2009}.   
Similarly, IBM Watson would have lost to elite Jeopardy
contestants if it waited too long to answer~\cite{lenchner-2011,ferrucci-aaai-2010}.  
However, OLDI and traditional services differ during query
execution.  Traditional services use structured databases to
retrieve answers, but OLDI services use loosely structured
or unstructured data.  Extracting answers from loosely  
structured data can be complicated.  Consider the
OpenEphyra question answering
system~\cite{openephyra-webpage}.  Each query execution 
reduces text documents to a few phrases by finding noun-verb
answer templates within sentences.

OLDI services use large and growing data to improve the
quality of their answers, but large data also increases
processing demands.  To keep response time low, OLDI query   
executions are parallelized across distributed software
components.  At Microsoft Bing, query execution invokes
over 100--1000 components in parallel~\cite{jalaparti-sigcomm-2013}. 
Each component contributes intermediate data that could improve
answers. However, some query executions suffer from slow
running components that take too long to complete. 
Since fast response time is essential, 
OLDI query executions cannot wait for slow components.
Instead, they compute answers with whatever data is
available within response time constraints.

OLDI services answer queries quickly even though performance
anomalies, failed hardware and skewed partitioning schemes
slow down some parallel components.  However, eliding data
from slow components could degrade answer
quality~\cite{ren-icac-2013,falsett2004limitation}. 
{In this paper,  answer quality is the similarity between answers
  produced with and without data from slow components.
}
Queries achieve high answer quality when their
execution does not suffer from slow components or when
slow components do not affect answers.  Low answer quality
means that slow, elided components have important
contributions that would affect final answers significantly.
Prior work has shown the virtue of adaptive resource
management with regard to response time.
Adaptive management could also help OLDI services manage
answer quality.  For example, adaptive admission control
could stabilize answer quality in the face of time-varying
arrival rates.  

\begin{figure}[t]
  \centering
  \includegraphics[width=3.0in]{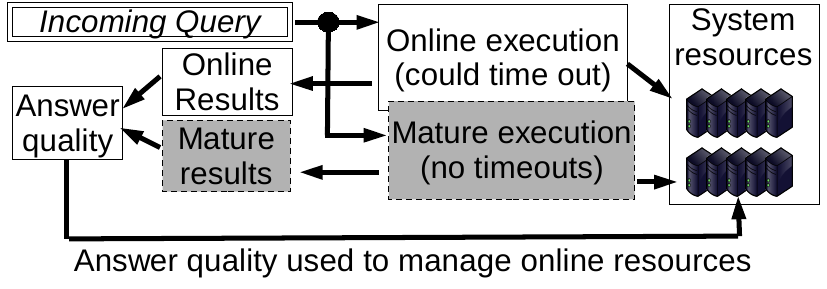}
  \caption{\MYcaptionfontsize Steps to measure answer quality
    online.  Mature and online executions may overlap. }
  \label{fig:draw-contributions}
  \vspace{-0.1in}
\end{figure}

Answer quality is hard to measure online because it requires
2 query executions. 
\MYfigureref{fig:draw-contributions} depicts the process of
computing answer quality.
First, an online execution provides answers within response
time constraints by eliding data from slow components. 
Second, a {\em mature execution} provides mature answers
by waiting for all components before producing answers.
Finally, a service-specific similarity function computes
answer quality.  This paper uses true positive rate as the
similarity function, but other functions are permissible,
e.g., normalized discounted cumulative gain~\cite{manning-2008}.
 
{\em We present Ubora\footnote{Ubora means {\it quality} in
    Swahili.}, an approach to speed up mature executions. }  
Our key insight is that mature and online executions invoke
many components with the same parameters.  Memoization can
speed up mature executions, i.e., a mature execution can  
complete faster by reusing  data from its
corresponding online execution instead of re-invoking
components. 

When a query arrives, Ubora conducts a normal online query
execution except it records intermediate data provided by
each component, including data from slow components that
were elided from online answers.  After the slow
components finish, Ubora computes {\em mature answers} using
data recorded during and after the online execution.  
Implementing memoization for multi-component OLDI
services presents systems challenges.
First, OLDI components span multiple platforms.   It is
challenging to coordinate mature and online executions
across components without changing application-level source code.   
Ubora manages mature and online operating context.  During
mature executions, it uses network redirection to replay
intermediate data stored in a shared key-value store.
Second, memoization speeds up computationally intensive
components but its increased bandwidth usage can also cause
slowdown for some components.  Ubora provides flexible
settings for memoization, allowing each component to turn
off memoization.  We use offline profiling to
determine which components benefit from memoization.

We have evaluated \systemname on Apache Lucene with
Wikipedia data, OpenEphyra with New York Times data, EasyRec
recommendation engine with Netflix data and Hadoop/Yarn with
BigBench data.   To be sure, Ubora's systems-level
implementation is able to support these applications without
modification to their source code.  We compared Ubora to a
competing approach that eschews transparency for
efficiency.  Specifically, this competing approach changes
application source code to tag query executions with custom
online and mature timeout settings.  A query executes until
its online timeouts trigger, it returns an answer and then
resumes for mature execution. 
Ubora completes mature executions nearly as quickly as this
approach with slowdown ranging from 8--16\%.
We also compared Ubora to an alternative approach that does
not require changing application source code.  
In this approach, each component's local timeout settings
are extended for a short period of time.  
Ubora finishes mature executions 7X faster.
Finally, Ubora slows down normal, online
query executions by less than 7\%.   

We also used Ubora to guide adaptive admission control.
We adaptively shed low priority queries to our Apache Lucene
and EasyRec systems. The goal was to maintain high answer
quality for high priority queries.   Ubora provided answer
quality measurements quickly enough to detect shifts in the
arrival rate and query mix.  Specifically, Ubora responded
quickly to changing arrival rates, keeping answer quality
above 90\% during most of the trace.  The other transparent
approach to measure answer quality, i.e., toggling timeouts,
produced mature executions too slowly. This
approach allowed answer quality to fall below 90\% 12X much more
often than Ubora.  We also used component
timeouts as a proxy for answer
quality~\cite{jalaparti-sigcomm-2013}.  This metric is
available after online executions without conducting
additional mature executions.  As a result, it has much lower
overhead.  However, component timeouts are a conservative
approximation of answer quality because they do not assess
the effect of timeouts on answers.  While achieving the same
answer quality on high priority queries, Ubora-driven
admission control processed 37\% more low priority queries
than admission control powered by component timeouts.

This paper is organized as follows.  We describe the
structure of OLDI services in
\MYsectionref{sect:workflows}.  We present Ubora in
\MYsectionref{sect:design}.   \MYsectionref{sect:ubora}
presents our implementation of query context tracking and
profiling for memoization.  In {\MYsectionref{sect:setup}},
we measure Ubora's performance using a wide range of OLDI
benchmarks.
In \MYsectionref{sect:onlinemanagement}, we show that
Ubora computes answer quality quickly enough to guide online
load shedding.

\section{Background on OLDI Services}
\label{sect:workflows}

Query executions differ fundamentally between online
data-intensive (OLDI) and traditional Internet services.  
In traditional services, query executions process data
retrieved from well structured databases, often via SQL. 
Correct query executions produce answers with well defined
structure, i.e., answers are provably right or wrong.
In contrast, OLDI queries execute on unstructured data.
They produce answers by discovering correlations within
data. OLDI services produce good answers if they process data
relevant to query parameters. 

Large datasets improve the quality of OLDI answers.  
For example, IBM Watson parsed 4TB of mostly public-domain
data~\cite{ferrucci-aaai-2010}.  One of Watson's data
sources, Wikipedia, grew 116X from 2004--2011~\cite{wikigrowth}. 
However, it is challenging to analyze a large dataset within
strict response time limits.  This section provides
background on the software structure of OLDI services that
enables the following: 
\begin{MYenumwide}
\item Parallelized query executions for high throughput,
\item Returning online answers based on partial, best-effort
  data to prevent slow software components from delaying response time.
\end{MYenumwide}

\begin{figure}[t]
  \centering
  \includegraphics[scale=0.21]{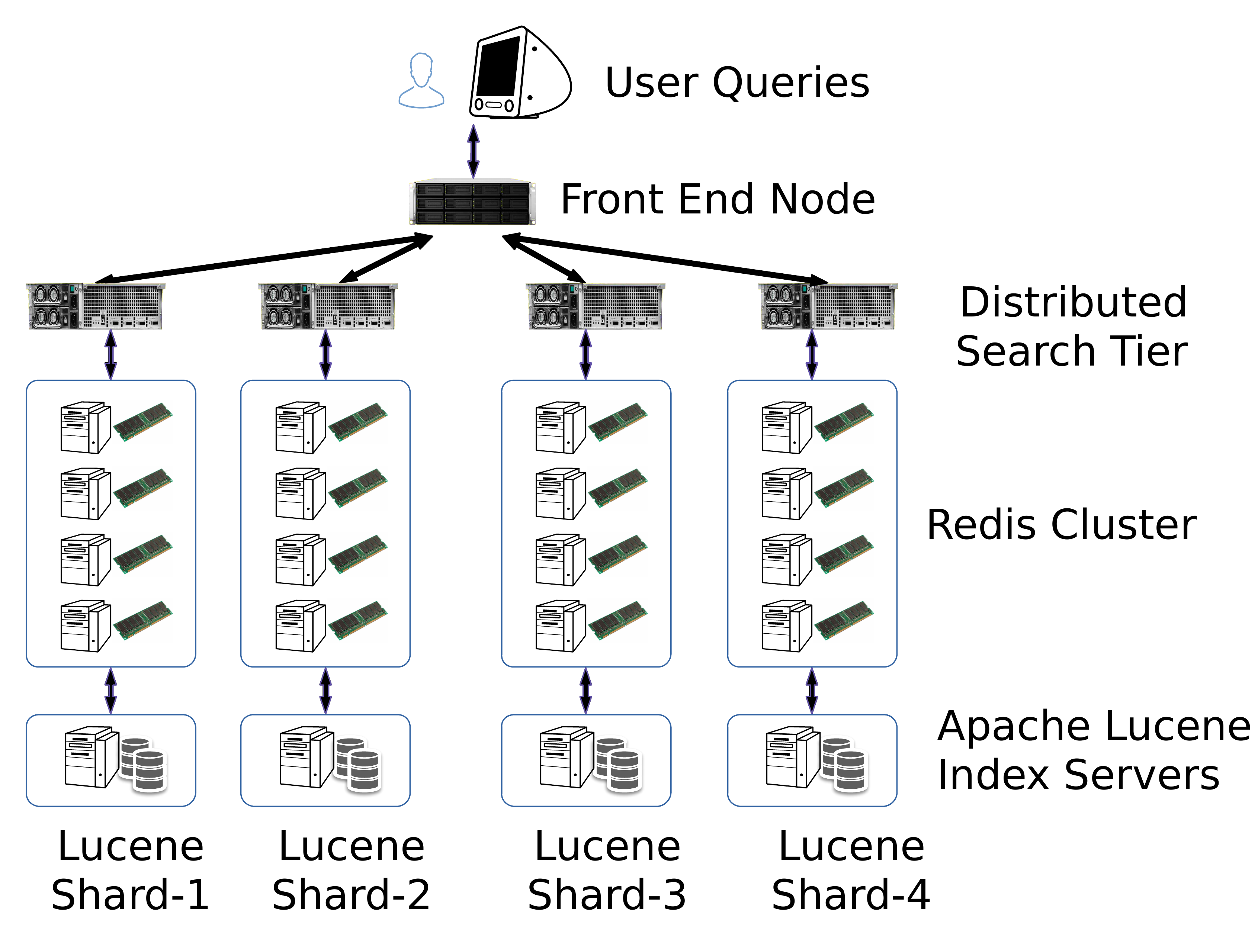}
  \caption{\MYcaptionfontsize Execution of a single query in Apache Lucene. 
    Adjacent paths reflect parallel execution across data partitions.}
  \label{fig:draw-solrworkflow}
\end{figure}

{\noindent \bf Parallelized Query Execution:}  
\MYfigureref{fig:draw-solrworkflow} depicts a query execution 
in an Apache Lucene system, a widely used open-source
information retrieval library~\cite{lucid-2008}. 
Query execution invokes 25 software components.
Components in adjacent columns can execute in parallel.
A front-end node manages network connections with
clients, sorts results from nodes running Distributed Search logic and produces
final answers. Distributed Search parses the query, requests a wide
range of relevant data from storage nodes, and collects data returned within a given
timeout.  Data is retrieved from either 1) an in-memory Redis cluster that caches a subset of index
entries and documents for a Lucene Index Server or 2) the Lucene Index Server
itself, which stores the entire index and data on relatively
slow disks.

The Lucene system in \MYfigureref{fig:draw-solrworkflow}
indexes 23.4 million Wikipedia and NY Times documents (pages
+ revisions)  produced between 2001 and 2013.   
It parallelizes query execution via data parallelism, i.e.,
the Lucene Index Servers partition the index across
multiple nodes.  Each parallel sub-execution (i.e., a
vertical column) computes intermediate data based on its
underlying partition.  Intermediate data is combined to
produce a query response.  

OLDI services also parallelize query executions via partial
redundancy.  In this approach, sub-executions compute
intermediate data from overlapping data partitions.  The
query execution weights answers based on the degree of
overlap and aggregate data processing per partition. 
Consider a product recommendation service.  Its query
execution may spawn two parallel sub-executions.  The first
finds relevant products from orders completed in the last
hour.  The second considers the last 3 days.  The service
prefers the product recommended by the larger (3-day)
sub-execution.  However, if the preferred recommendation is
unavailable or otherwise degraded, the results from the smaller parallel
sub-execution help.

\begin{figure}[t]
  \centering
  \includegraphics[width=2.0in]{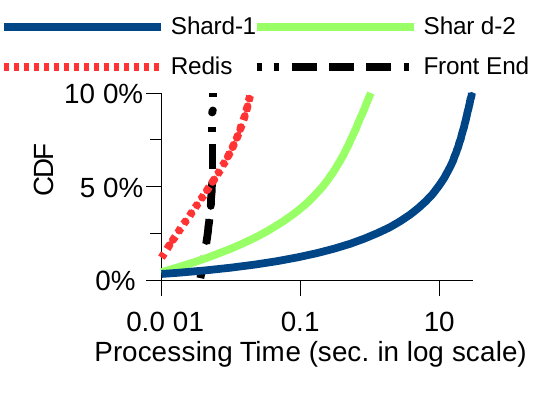}
  \caption{\MYcaptionfontsize  OLDI components exhibit
    diverse processing times. }
  \label{fig:xcomponents-yrunningtimes}
\end{figure}

{\noindent \bf Online Answers Are Best Effort:}
In traditional Internet services, query execution invokes
software components sequentially.  Their response time depends on
aggregate processing times of all components.  In contrast,
online data-intensive query executions invoke components in
parallel. The processing time of the slowest components
determines response time.
\MYfigureref{fig:xcomponents-yrunningtimes} quantifies
component processing times in our Apache Lucene system.
The query workload from Google Trends and hardware
details are provided in \MYsectionref{sect:setup}.
Processing times vary significantly from query
to query. Note, the X-axis is shown in log scale.  
Lucene Index servers can take several seconds on some
queries even though their typical processing times are much
faster. Further, processing time is not uniform across shards.  
For example, a query for ``William Shakespeare'' transferred
138KB from the shard 4 execution path but only 1KB from the shard 1 
execution path.
Shard 4 hosted more content related to this query even
though the data was partitioned randomly.

Many OLDI services prevent slow components from delaying response
time by returning answers prematurely---before slow
components finish.  Specifically, query executions trigger
timeouts on slow components and produce answers that exclude
some intermediate data.  Timeouts effectively control
response time.  In our Apache system, we set a 2 second and a
4 second timeout in our front-end component.  Average
response time fell.  Also, third quartile response times
were consistently close to median times, showing that
timeouts also reduced variance.
Unfortunately, query executions that trigger timeouts
use less data to compute answers.  This degrades answer
quality.  For data-parallel queries answer quality degrades
if the elided data is relevant to query parameters.  

\begin{figure}[t]
\centering
  \includegraphics[width=3.1in]{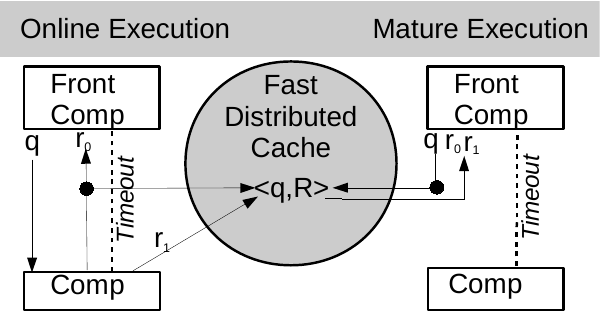}
  \caption{\MYcaptionfontsize Memoization in Ubora.  Arrows reflect messages
    in execution order (left to right). }
  \label{fig:draw-rcr}
\end{figure}

\section{Design}
\label{sect:design}

Ubora measures the answer quality of online query
executions.  By design, it directly computes answer quality
by comparing answers produced with and without timeouts.
It uses existing online resources and employs memoization to
speed up query executions.  

\MYfigureref{fig:draw-rcr} depicts memoization in Ubora.
During online query execution, Ubora records inter-component
communication.  It allows only front-end components to time
out.  Components invoked by parallel sub-executions complete
in the background.  
As shown on the left side of \MYfigureref{fig:draw-rcr}, 
without Ubora, the front-end component invokes a component
with query $q$, receives message $r_0$ and then times out.
The front-end component then triggers a timeout for the
invoked component, stopping its execution prematurely. 
Ubora blocks the trigger from the front-end component,
allowing the invoked component to complete a mature execution.
It records output messages before and {\em after} the
front-end times out, in this case $r_0 + r_1$.  These
messages are cached in fast, in-memory stores.  

With Ubora, front-end components still answer online queries
within strict response time limits.  As shown in
\MYfigureref{fig:draw-rcr}, the front-end component 
uses $r_0$ to produce an {\em online answer}.  After all
sub-executions for a query complete, Ubora 
re-executes the front-end, as if a new query arrived.
However, during this re-execution, Ubora intercepts messages
to other components and serves from cache (i.e., memoization).
The cache delivers messages with minimal processing or disk delays.  
During this mature execution, the front-end uses both  $r_0 + r_1$ to
produce a {\em mature answer}.  

\section{Implementation}
\label{sect:ubora}

This section discusses the implementation of Ubora.
First, we describe axiomatic choices, e.g., the user
interface, target users and prerequisite infrastructure.
Second, we discuss the impact of
operating system support on the implementation of memoization.
Finally, we provide details about our implementation, including our approach to
determine which components constitute a front-end.  

\subsection{Interface and Users}
Ubora targets system managers.  It runs on a cluster of compute nodes.
Each node runs a networked operating system.  Many software components can map
to each node, but each component runs on just 1 node.  To be
clear, a software component is a running binary that accepts
invocations over the network.  Software components have unique network
addresses, e.g., IP address and TCP port.

System managers understand the query execution paths in their system (e.g.,
as depicted in~\MYfigureref{fig:draw-solrworkflow}).  They classify each
component as front- or back-end.  Front components receive
queries, record inter-component messages and produce online
and mature answers.  They are re-executed to get mature
answers.  Back-end components propagate query context,
record messages, and do not time out for sampled queries.
\MYfigureref{fig:draw-solrworkflow} labels the front-end
component.  The search tier, Redis and/or Lucene could be
front-end or back-end components.  

Ubora is started from the command line.  Two shell scripts, {\em
  startOnBack and startOnFront}, are run from a front component.
Managers can configure a number of parameters before starting Ubora, shown
in Listing~\ref{listing:config}.
The number of mature executions to produce per unit time controls the query
sampling rate.  When new queries arrive at front end TCP ports, a query sampler
randomly decides how to execute the query.  Sampled queries
are executed under the {\em record mode} context shown on
the left side of \MYfigureref{fig:draw-rcr}.  Queries not
sampled are executed normally without intervention from Ubora.
Record timeout duration sets the upper bound on processing time for
a back-end component's mature execution. Propagate timeout
is used to set the upper bound on time to scan for newly
contacted components to propagate the execution context.
To get mature answers the query execution context is called
{\em replay mode}.  
Finally, the callback function used to
compute answer quality is service specific.  The default is True Positive Rate.

\subsection{Impact of Operating System Support}
A key goal was to make \systemname as transparent as possible.
Here, transparent means that 1) it should work with existing middleware and 
operating systems without changing them and 2) it should have small effects on
response times for online queries. 
Transparency is hard to achieve, because \systemname must manage record and
replay modes without changing the interaction between software components.
In other words, the execution context of a query must pass between components unobtrusively.
Some operating systems already support execution contexts.  Therefore, we present 
two designs. The first design targets these operating systems.  The second design targets commodity operating
systems.  Our designs exploit the following features of memoization:
\begin{MYenumwide}
\item {\em Queries produce valid output under record, replay, and normal modes.}
  This property is achieved by maintaining a shadow connection to the invoked
  component during replay.  Cache misses trigger direct communication with
  invoked components.  As a result, 
  replay, normal, and record modes have access to full data.
\item {\em Back-end components use \MYnote{a lot} more resources during record mode than they
  use during normal online execution because timeouts are disabled.}
\end{MYenumwide}

\lstset{basicstyle=\small\selectfont\ttfamily,breaklines=true}
\lstset{
frame = single, 
language=Pascal, 
}

{\small
\centering
\begin{lstlisting}[linewidth=2.9in,float,captionpos=b,caption={\small
      Ubora's YAML Configuration.},label={listing:config}]
IPAddresses
-  front: 10.243.2.*:80
-  back: 10.244.2.*; 10.245.2.*:1064

samples: 8 per minute
recordTimeout: 15 seconds
propagateTimeout: 0.1 seconds
answerQualityFunction: default 
\end{lstlisting}
}

{\bf \noindent Design with OS Managed Contexts:}
Some operating systems track execution context by annotating
network messages and thread-local memory with context and ID.  
Dapper~\cite{sigelman-tr-2010} instruments Google's threading libraries, Power
Containers~\cite{shen-asplos-2012} tracks context switches between Linux
processes and annotates TCP messages and Xtrace~\cite{fonseca-nsdi-2007}
instruments networked middleware.

OS-managed execution context simplifies our design.
\systemname intercepts messages between components, acting as a middle box.  
Before delivering messages that initiate remote procedures, \systemname checks
query ID and context and configures memoization-related
context (i.e., record or replay mode).  The same
checks are performed on context switches.  
During record mode, when a component initiates a remote
invocation, we use the message and query id as a key in the
cache.  Subsequent component interactions comprise the value
associated with the key---provided the query context and 
ID are matched.   We split the value and form a new key when
the invoking component sends another message.    

In replay mode, when an invocation message is intercepted, 
the message is used to look up values in the cache.  On hits, the cache returns
all values that are associated with the message. The cache results are turned into
properly formatted messages to transparently provide 
the illusion of RPC. On misses, the message is delivered to
the destination component as described above.

{\bf \noindent Design without OS Support: }
Most vanilla operating systems do not track execution context.
Without such support, it is challenging to distinguish remote procedure calls
between concurrent queries.   However, Ubora's memoization permits
imperfect context management because record, replay and normal modes yield valid
output.  This feature allows us to execute concurrent queries under the same
context.
First, we describe a simple but broken idea that is highly transparent, and then we present an
empirical insight that allows us to improve this design without sacrificing transparency.

In this simple idea, each component manages its current, global execution context
that is applied to all concurrent queries.  Also, it manages a context id that
distinguishes concurrent record contexts.  Ubora intercepts
messages between components.  When a component initiates a
remote invocation in record mode, the message and context id are
used to create a key.  For the duration of record mode,
inter-component messages are recorded as values for the key.   
If the context indicates replay mode, the message and context id are used to retrieve
values from cache.

This simple idea is broken because all messages from the
invoked component are recorded and cached, including
concurrent messages from different queries.
In replay mode, those messages can cause wrong output.  Our key insight is that 
record mode should use replies from the invoked component only if they are from the same 
TCP connection as the initiating TCP connection.  
The approach works well as long as TCP connections are not shared by
concurrent queries.  Widely used paradigms like TCP connection
pooling and thread pooling are ideal for use with Ubora.  We studied the source
code of 15 widely used open source components including: JBoss, LDAP,
Terracotta, Thrift and Apache Solr.  Only 2 (13\%) of these platforms
multiplexed concurrent queries across the same connection.  This suggests that
our transparent design can be applied across a wide range of services.  We
confirm this in \MYsectionref{sect:overhead}.

Next we describe how to propagate request context, which is
necessary when the operating system does not support
execution contexts.  On a front component, 
we wait for queries to arrive on a designated TCP port.  If a query is selected
for mature execution, we change the front component context from normal to
record and create a context id.  Before sending any TCP message, we extract the
destination component. If the destination has not been contacted since record
mode was set, we send a UDP control message that tells that component to enter
record mode and forwards the proper context id.  Then we send the original
message.  Note, UDP messages can fail or arrive out of order.  This causes the
mature execution to fail.  However, we accept lower throughput  (i.e., mature
executions per query) when this happens
to avoid increased latency from TCP roundtrips.  Middle components propagate
state in the same way.  Each component maintains its own local timers.  After
a propagation timeout is reached, the context id is not forwarded anymore.  After the
record timeout is reached, each component reverts back to normal mode independently.
We require front components to wait slightly longer than record timeout to ensure the system has returned to normal.

{\bf \noindent Reducing Bandwidth Needs:}
Ubora reduces bandwidth required for context propagation. First, Ubora propagates context to
only components used during online execution.  Second, Ubora does not use 
bandwidth to return components to normal mode, only sending UDP messages when 
necessary to enable record or replay mode. Timeouts local to each component also 
increase robustness to network partitions and congestion, ensuring that components 
always return to normal mode.

\subsection{ Determining Front-End Components }
Thus far, we have described the front-end as the software
component at which queries initiate.  Its internal timeout
ensures fast response time, even as components that it
invokes continue to execute in the background.  To produce
an online answer, the front-end must complete its
execution.  Ubora re-executes the front-end to get mature
answers.  Ubora can not apply memoization to the front-end
component.  

At first glance, re-execution seems slower than
memoization.  However, as shown in
\MYfigureref{fig:xcomponents-yrunningtimes}, many components
execute quickly.  In some cases, execution is faster than
transferring intermediate data to the key-value store.  
Our implementation allows for a third class of component:
middle components.  Like front-end components, middle
components are allowed to time out.  They are re-executed in
replay mode without memoization.  Unlike front-end
components, middle components do not initiate queries.  In
\MYfigureref{fig:draw-solrworkflow}, Distributed Search or
Redis components could be labeled middle components.  

Given a trace of representative queries, Ubora determines
which components to memoize by systematically measuring
throughput with different combinations of front-,
middle- and back-end components.  We do the same to
determine the best sampling rate.

\subsection{Prototype}

We implemented transparent context tracking as described
above for the Linux 3.1 operating system.  
The implementation is installed as user-level package and requires the Linux
Netfilter library to intercept and reroute TCP messages.  It uses IPQueue to
trigger context management processes.  It assumes components communicate through 
remote procedure calls (RPC) implemented on TCP and that an IP address  and TCP
port uniquely identify each component.  It also assumes timeouts are triggered
by the RPC caller externally---not internally by the callee.
It extends timeouts transparently by blocking FIN packets sent to the callee and
spoofing ACKs to the caller.  Messages from the callee that arrive
after a blocked FIN are cached but not delivered to the caller.
For workloads that use connection pooling, we block application-specific
termination payloads.  Service managers can specify this in the configuration file.  

We use a distributed Redis cache for in-memory key value storage.  Redis allows
us to set a maximum memory footprint per node.  The aggregate memory across all
nodes must exceed the footprint of a query.  Our default setting is 1 GB.
Also, Redis can run as a user-level process even if another Redis instance runs
concurrently, providing high transparency.

We want to minimize the overhead in terms of response time and cache miss 
rate.  Each key value pair expires after a set amount of time.  Assuming a set
request rate, cache capacity will stabilize over time.  A small 
amount of state is kept in local in-memory storage on the \systemname control
unit node (a front node).  Such state includes sampled queries, online and
mature results and answer quality computations.

\section{Experimental Evaluation}
\label{sect:setup}

\begin{table*}[tbp]
\begin{center}
\small

\begin{tabular}{|l|l|l|r|r|r|r|r|}
\hline
\textbf{Code-} & \textbf{Platform} & \textbf{Parallelism} &
\multicolumn{1}{l|}{\textbf{Parallel}} &
\multicolumn{1}{l|}{\textbf{Data}} &
\multicolumn{1}{l|}{\textbf{Nodes}} &
\multicolumn{1}{l|}{\textbf{Online resp.}} &
\multicolumn{1}{l|}{\textbf{Mature resp.}} \\ 
\textbf{name} & \textbf{} & \textbf{} &
\multicolumn{1}{l|}{\textbf{Paths}} &
\multicolumn{1}{l|}{\textbf{(GB)}} &
\multicolumn{1}{l|}{\textbf{}} &
\multicolumn{1}{l|}{\textbf{time (sec)}} &
\multicolumn{1}{l|}{\textbf{time (sec)}} \\ \hline
\textbf{YN.bdb} & Apache Yarn & Partial Rep. & 2 & 1 & 8 &
178.00 & 185.00 \\ 
\textbf{LC.lit} & Lucene & Data & 1 & 4 & 4 & 1.00 & 1.22\\ 
\textbf{LC.wik} & Lucene & Data & 4 & 128 & 31 & 3.00 & 8.97\\ 
\textbf{LC.big} & Lucene & Data & 4 & 4096 & 31 & 5.00 &
23.52 \\ 
\textbf{ER.fst} & EasyRec & Partial Rep. & 2 & 2 & 7 & 0.50
& 0.60 \\ 
\textbf{OE.jep} & OpenEphyra & Data & 4 & 4 & 8 & 3.00 &
23.43 \\ \hline
\end{tabular}

\end{center}
\caption{The OLDI workloads used to evaluate Ubora supported
  diverse data sizes and processing demands.}
\label{tab:workloads}
\end{table*}

In this section, we compare Ubora to alternative designs and
implementations across a wide range of OLDI workloads.
First, we discuss the chosen metrics of merit.  Then, we
describe the competing designs and implementations.  Then,
we present the software and hardware architecture for the
OLDI services used.  Finally, we present experimental
results. 

\subsection{Metrics of Merit}
Ubora speeds up mature query executions needed to compute
answer quality.  The research challenge is to complete
mature query executions while processing other queries
online at the same time.  The primary metric used to
evaluate Ubora's performance ({\em throughput}) is mature
executions completed per online execution.   

Mature executions use resources otherwise reserved for
online query executions, slowing down response times.
Online queries that Ubora does not select for mature
execution (i.e., unsampled queries) are slowed down by
queuing delays.  We report {\em slowdown} as the relative
increase in response time.   In addition to queuing delay,
online queries sampled for mature execution are also slowed
down by context tracking and memoization. 

Finally, we used {\em true positive rate}, i.e., the percentage 
of mature results represented in online results, to compute answer quality.

\subsection{Competing Designs and Implementations}
Ubora achieves several axiomatic design goals.
Specifically, it (1) speeds up mature executions via
memoization, (2) uses a systems approach that works for
a wide range of OLDI services, (3) supports adjustable
query sampling rate and (4) implements optimizations that
reduce network bandwidth.  Collectively, these goals
make Ubora usable.  Our evaluation compares competing
designs that sacrifice one or more of these goals.  They are
listed below with an associated codename that will be used
to reference them in the rest of the paper.
\begin{MYlistwide}
\item {\em Ubora} implements our full design and
  implementation.  The sampling rate is set to maximize
  mature query executions per online query.
\item {\em Ubora-LowSamples} implements our full design and
  implementation, but lowers the sampling rate to reduce slowdown.
\item {\em Ubora-NoOpt} disables Ubora's optimizations.
  Specifically, this implementation disables node-local
  timeouts that reduce network bandwidth.
\item {\em Query tagging and caching } essentially
  implements Ubora at the application level.  Here, we
  implement context tracking by changing the OLDI service's
  source code so that each query accepts a timeout
  parameter.  Further, we set up a query cache to reuse
  computation from online execution.  This approach is
  efficient but requires invasive changes.
\item {\em Query tagging } implements context tracking at
  the application level but disables memoization.
\item {\em Timeout toggling} eschews both context tracking and
  memoization.  This implementation increases each component's
  global timeout settings by 4X for mature executions.  All
  concurrent query executions also have increased timeout settings.
  This is non-invasive because most OLDI components support
  configurable timeout policies.  However, extending
  timeouts for all queries is costly.
\end{MYlistwide}

\begin{figure*}[t!]
  \includegraphics[width=6.0in]{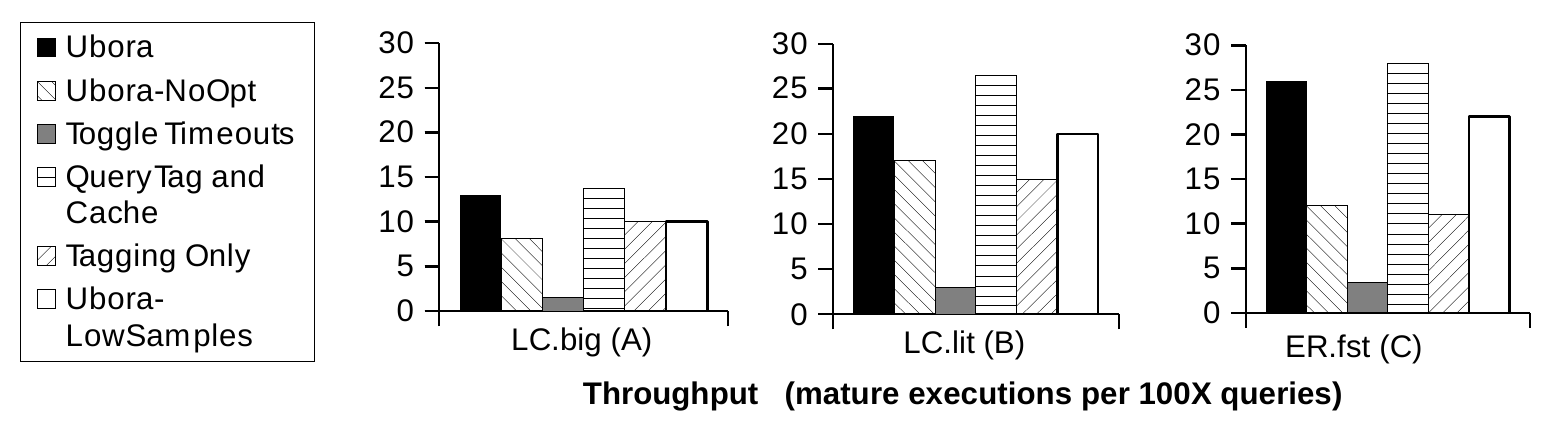}

  \hrulefill
  \caption{\small {\bf Experimental results: }
    Ubora achieves greater throughput than competing
    systems-level approaches.  It performs nearly as well as
    invasive application-level approaches (within 16\%).  }

  \label{fig:xsamplerate-ythroughput}

\end{figure*}

\subsection{OLDI Services}
\MYtableref{tab:workloads} describes each OLDI service used
in our evaluation.  In the rest of this paper, we will
refer to these services using their codename.  
The setup shown in \MYfigureref{fig:draw-solrworkflow} depicts LC.all, a 31
node cluster that supports 16 GB DRAM cache per TB stored on disk.  Each
component runs on a dedicated node comparable to an EC2 medium instance, providing access to an Intel
Xeon E5-2670 VCPU, 4GB DRAM, 30 GB local storage and (up to) 2 TB block storage.

\begin{MYlistwide}
\item {\em YN.bdb} uses Hadoop/Yarn for sentiment analysis.
  Specifically, it runs query 18 in BigBench, a data
  analytics benchmark~\cite{ghazal-sigmod-2013}. 
  Each query spawns two parallel executions.  
  The first sub-execution extracts sentiments from customer
  reviews over 2 months.  The second covers 9 months.  
  The 9-month execution returns the correct answer, but the
  1-month answer is used after a 3-minute timeout.
  Each sub-execution flushes prior cached data in HDFS,
  restores a directory structures and compresses its
  output.  As a result, query execution takes minutes, even 
  though customer reviews are smaller than 1 GB.
  The average response time without timeout is 3 minutes.
  44\% of queries get the 9-month answer within timeout. 
  We mainly include YN.bdb to show that Ubora can capture
  answer quality for longer running services too.
\item {\em LC.big}, {\em LC.wik} and {\em LC.lit} use Apache
  Lucene for bag-of-words search.  All of these workloads
  replay popular query traces taken from Google Trends. LC.lit
  hosts 4GB of news    articles and books on cluster with
  16GB DRAM.  LC.lit   implements one parallel path shown in
  \MYfigureref{fig:draw-solrworkflow}.  It returns the best
  answer produced within 1 second.  Without
  timeouts, the average response time is 1.22 seconds.  
  Over 83\% of LC.lit queries complete within the timeout.

  {\em LC.wik} hosts 128GB of data from Wikipedia, New York
  Times and TREC NLP~\cite{trec}.  After executing warm-up
  queries, the data mostly fits in memory.  We set a
  3-second timeout.  Without the timeout, response time 
  was 8.9 seconds.  39\% of the LC.wik queries
  complete within the timeout.   
  {\em LC.big} hosts 4TB.  Most queries access
  disk.  Average response time without timeout is 23
  seconds.  The timeout is 5-seconds.

\item {\em ER.fst} uses the EasyRec platform to recommend
  Netflix movies.  It compiles two recommendation databases
  from Netflix movie ratings~\cite{netflix-data}:
  A 256MB version and a 2GB version.  Each query provides a
  set of movie IDs that seed the recommendation engine.  
  The engine with more ratings normally takes longer to
  respond but provides better  recommendations.  
  Query execution times out after 500 milliseconds.  80\% of
  query executions produce the 2GB answer.  

\item {\em OE.jep} uses OpenEphyra, a question answering 
  framework~\cite{openephyra-webpage}.  OpenEphyra uses
  bag-of-words search to extract sentences in NLP data
  related to a query.  It then compares each sentence to a
  large catalog of noun-verb templates, looking for specific
  answers.  The workload is computationally intensive.  The
  average response time in our setup was 23 seconds.
  Motivated by the responses times for IBM Watson, we set a
  timeout of 3 seconds~\cite{ferrucci-aaai-2010}.
  Fewer than 15\% of queries completed within timeout.

\MYnote{analyzes this data to answer a
  trace of 1K questions related to recent events, sent to a front end running on
  NanoWeb~\cite{nanoweb-webpage}.  The front end aggregates results from
  parallel invocations of OpenEphyra, a question 
  answering framework~\cite{openephyra-webpage}. OpenEphyra pulls whole web
  pages from our MYSQL-based disk storage. It uses Lucene and Redis for indexing 
and data caching.}

\end{MYlistwide}

We set up a workload generator
that replayed trace workloads at a set arrival rate.  Our workload generator kept
CPU utilization between 15--35\%.  

\subsection{Results}
\label{sect:overhead}

{\bf \noindent Comparison to Competing Approaches:}
\MYfigureref{fig:xsamplerate-ythroughput} compares
competing approaches in terms of mature executions completed
per query.  Ubora-NoOpt reveals that node-local timeouts and
just-in-time query propagation collectively improve
throughput by 1.6X, 1.3X and 2.1X respectively.  ER.fst
has relatively fast response times which require messages to
turn off record and replay modes.  Node-local timeouts
reduce these costs.  Internal component communications in
LC.big and LC.wik also benefit from node-local timeouts.

Excluding Ubora, the other competing approach that can be
implemented for a wide range of services is toggling timeouts.
Unfortunately, this approach performs poorly, lowering
throughput by 7-8X.  
To explain this result, we use a concrete example of a
search for ``Mandy Moore'' in LC.big.  
First, we confirm that both Ubora and toggling timeouts
produce the same results.  They produce the same top-5 results and 90\%
of the top-20 results overlap.  Under 5-second timeout, the query times out
prematurely, outputting only 60\% of top-20 results.
Ubora completes mature executions faster because it maintains execution context.
This allows concurrent queries to use different timeout settings.   Queries
operating under normal timeouts free resources for the
mature execution.  Further,
per-component processing times vary within mature
executions (recall, \MYfigureref{fig:xcomponents-yrunningtimes}).  By maintaining execution context,  Ubora avoids
overusing system resources.
For the ``Mandy Moore'' query, Ubora's mature execution took
21 seconds in record mode and 4 seconds in replay mode. 
Conversely, under the toggle timeouts approach, service
times for all concurrent queries  increased by 4X, exceeding
system capacity and taking 589 seconds. 

We also compared Ubora, a systems level approach that
strives to transparently support a wide range of services,
to application-level approaches.  Application-level approaches
can track query context efficiently by tagging queries as
they traverse the system~\cite{fonseca-nsdi-2007}.
Specifically, we modified LC.big, LC.lit and ER.fst to pass
query context on each component interaction.  Further, we
implemented a query cache for targeted query
interactions~\cite{amza2005transparent,guo2013vcache,paiva2013autoplacer}.
Our cache uses the Ubora's mechanism for memoization but
tailors it to specific inter-component interactions and
context ids.  As such, our application-level approach is
expected to outperform Ubora, and it does.  However, Ubora
is competitive, achieving performance within 16\% on all
applications.  We also compared to a simple
application-level approach that disables query caching.
This approach shows that memoization improves throughput by
1.3X on LC.big, 1.7X on LC.lit and 2.5X on ER.fst.
The benefit provided by memoization is correlated with the
ratio of mature execution times to online execution times.
In ER.fst, mature executions are mostly repeating online executions.

\begin{figure*}[t!]
  \includegraphics[width=6.0in]{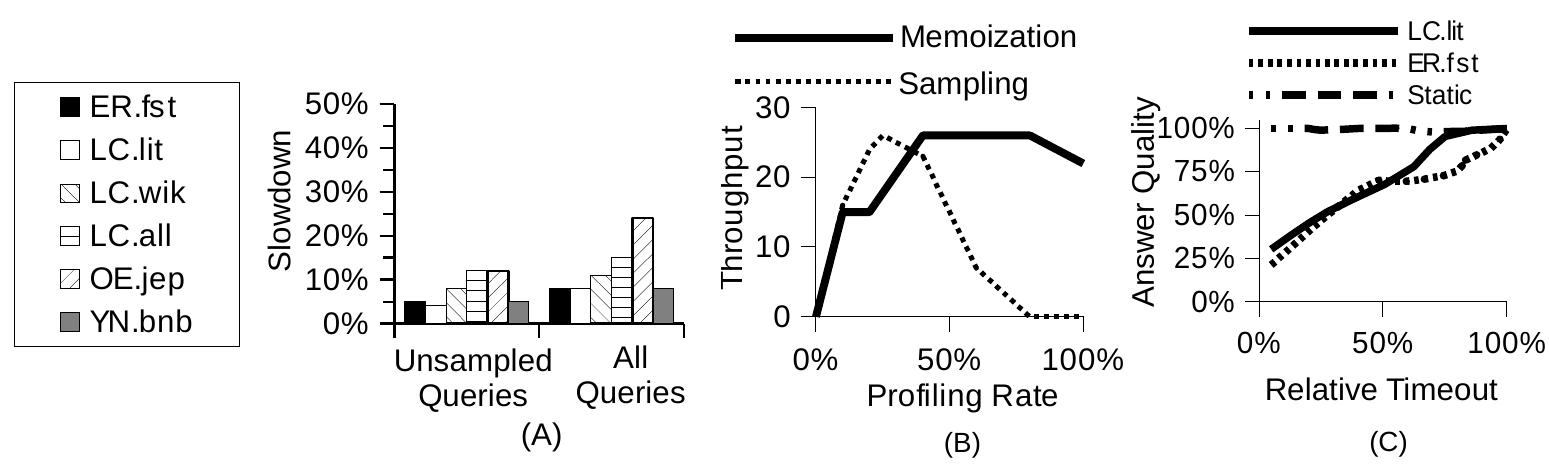}

  \hrulefill
  \caption{\small {\bf Experimental results: }   (a) Ubora
    delayed unsampled queries by 7\% on average. Sampled
    queries were slowed by 10\% on average.  (b) Profiling
    sampling and memoization options maximizes throughput.
    (c) Timeout settings have complex, application-specific
    affects on answer quality. }
  \label{fig:uborafiguresv3}
  \vspace{-0.2in}
\end{figure*}

{\bf \noindent Impact on Response Time:}
Ubora allows system managers to control the query sampling
rate.  As shown in
\MYfigureref{fig:xsamplerate-ythroughput}, a slight
reduction in the sampling rate can still achieve high
throughput.  However, this approach significantly reduces
Ubora's effect on response time.  
\MYfigureref{fig:uborafiguresv3}(A) shows the slowdown
caused by the Ubora-LowSamples approach across all tested
workloads.  By executing mature executions in the background
and staying within processing capacity, we achieve slowdown
below 13\% on all workloads for unsampled queries and below
10\% on 4 of 6 workloads for sampled queries.  OpenEphyra
and LC.all incur the largest overhead because just-in-time
context interposes on many inter-component interactions due
to cluster size.  For such workloads, OS-level context
tracking would improve response time for
sampled queries. 

{\bf \noindent Impact of Profiling:}
\MYfigureref{fig:uborafiguresv3}(B) studies our
approach to determine sampling rate and front-end components
(i.e., memoization).  We studied the ER.fst workload.  
Along the x-axis, we vary the sampling rate and the
percentage of components included as front-end of middle
components.  The y-axis shows the achieved throughput.  
For the ER.fst workload it is better to apply memoization to
many components.  The ideal sampling rate was 20\%.  

{\bf \noindent Studying Answer Quality:}
\MYfigureref{fig:uborafiguresv3}(C) shows answer quality
(i.e., the true positive rate) as we increase timeout
settings.  For LC.lit and ER.fst, we increase timeouts at
front-end components.  We also validate our results by
increasing timeouts in an unrelated component in ER.fst (Static).  We
observe that answer quality is stable in the static
setting.  Further, answer quality curves differ between
applications.  After timeout settings reach 600 milliseconds
for LC.lit and 300 milliseconds for ER.fst, the curves
diverge and answer quality increase slowly for ER.fst.
Finally, answer quality curves have 2 phases in LC.lit and 3
phases in ER.fst.  It is challenging to use timeouts to
predict answer quality.

\section{Online Management}
\label{sect:onlinemanagement}

OLDI services are provisioned to provide target response times.  In addition to 
classic metrics like response time, these services could use answer quality to 
manage resources.  We show here that Ubora enables better resource management through answer quality.

\begin{figure}[t!]
    \centering
    \small
    \includegraphics[width=3.3in]{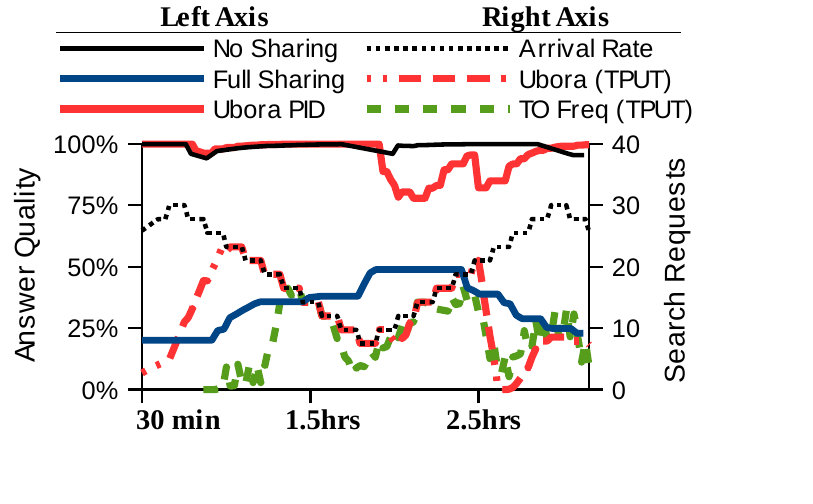}
    \vspace{-0.2in}
    \caption{Ubora  enables online load shedding.}
    \label{fig:xtime-y1answerquality-y2loadshed}
\end{figure}

{\noindent \bf Control Theory with Answer Quality:}
We studied load shedding on the LC.big workload.  Using diurnal traces from previous
studies~\cite{stewart-eurosys-2007}, we issued two classes of queries: high and
low priority.   The queries were directed to two different TCP ports.  
At the peak workload, low and high priority arrival rates saturate
system resources (i.e., utilization is 90\%).
\MYfigureref{fig:xtime-y1answerquality-y2loadshed} shows the {\em Arrival Rate} over
time (on the right axis).  At the 45 minute and 2 hour mark, the query mix
shifts toward multiple word queries that take longer to process fully.

We used \systemname to track answer quality for high
priority queries.  
When quality dipped, we increased the load shedding rate on low priority
queries.  Specifically, we used a proportional-integral-derivative (PID)
controller.  Every 100 requests, we computed answer quality
from 20 sampled queries (20\% sample rate). 

The left axis of \MYfigureref{fig:xtime-y1answerquality-y2loadshed} shows answer
quality of competing load shedding approaches.  When all low priority queries are
shed, the {\em No Sharing} approach maintains answer quality above 90\%
throughout the trace.  When shedding is disabled, the {\em Full Sharing}
approach sees answer quality drop as low as 20\%, corresponding with peak
arrival rates.  The PID controller powered by Ubora manages the shed rate well, keeping
answer quality above 90\% in over 90\% of the trace.  It maintains throughput
({\em Ubora TPUT})
of almost 60\% of low priority queries (shown on the right axis).  

The state of the art for online management in OLDI services is to use proxies
for the answer quality metric.  Metrics like the frequency of timeouts provide a
rough indication of answer quality and are easier to compute online~\cite{jalaparti-sigcomm-2013}.  
For comparison, we implemented a PID controller that used frequency of timeouts
instead of answer quality.  We tuned the controller to achieve answer quality
similar to the controller based on answer quality.  
However, timeout frequency is a conservative indicator
of answer quality for Lucene workloads.  It assumes that partial results caused by timeouts are
dissimilar to mature results.  \MYfigureref{fig:xtime-y1answerquality-y2loadshed}
also shows that the controller based on timeout frequency ({\em TO Freq}) sheds requests too
aggressively.  Our approach improved
throughput on low priority queries by 37\%.

{\bf \noindent Sampling Rate and Representativeness:}
Ubora allows reducing the overhead of mature executions by
sampling online executions.  This lowers mature results per
query, but how many mature results are needed for online
management? \MYnote{ \MYtableref{tab:representativeness} shows the
effect of lower sampling rates on the accuracy of answer
quality measurements and on the outcome of adaptive load
shedding.}  We observed that sampling 5\% of online queries 
significantly increased outlier errors on answer quality,
but our adaptive load shedding remained effective---it still
achieved over 90\% quality over 90\% of 
the trace.  In contrast, a 2\% sampling produced many quality
violations.

\section{Related Work}
\label{sect:relatedwork}

Ubora focuses on online systems, which trade answer quality for fast response times.  Zilberstein first characterized similar applications as anytime algorithms~\cite{Zaimag96}. Like the online, data-intensive workloads used with Ubora, anytime algorithms increase in quality of result as they increase in computation time.  Zilberstein uses three metrics, certainty, accuracy, and specificity; accuracy is closest to our answer quality metric, but does not indicate how the exact answer is to be reached for comparison. His work indicates that anytime algorithms should have measurable quality, monotonically increase in quality as computation time increases, provide diminishing returns, and produce a correct answer when interrupted.  Properties of anytime algorithms that Ubora does not need to work correctly include recognizable quality and being able to suspend and resume application processing at any point in time. Recognizable quality refers to ability to determine the optimal answer in constant time for online processing; Ubora decreases the total time to acquire a mature execution by reducing data access times, but does not interfere with the processing of this data by software components.

Recent work has focused on introducing approximation into existing systems in order to increase performance~\cite{goiri2015approxhadoop,jeon-eurosys-2013,jalaparti-sigcomm-2013}.
Goiri et al. created ApproxHadoop~\cite{goiri2015approxhadoop} to integrate sampling of input data, user-defined approximate code versions, and execution of only a subset of tasks into Hadoop. The authors allow users to set error bounds within a confidence interval, set a specific data sampling ratio, or specify the percentage of tasks to drop in order to increase performance. ApproxHadoop uses extreme value theory and multi-stage sampling theory to compute error bounds and confidence intervals depending on type of aggregation query.  Our work with Ubora uses mature executions to compute answer quality, and enables users to similarly manage resources based on the online answer quality trace.
Jeon et al. worked on intra-query parallel index search within a Bing server~\cite{jeon-eurosys-2013}. Sequential search may terminate early on a server if the processing of the ranked documents goes below a certain relevance. Since parallel search over the same index will generally result in more processing per query, the authors reduce this wasted work by keeping the order in which documents are processed sequential.  While this is not necessary under low load, higher loads are more impacted by wasted work. The authors adaptively change the amount of parallelism per query based on the current system load.
Jalaparti et al. introduced Kwiken, their optimization framework for lowering tail latency in Bing~\cite{jalaparti-sigcomm-2013}. Techniques the authors used within Kwiken include allowing return of incomplete results, reissuing queries that lag on different servers, and increasing apportioned resources. They calculate incompleteness as utility loss based on whether the answer returned contains the highest ranked document for certain stages, and in other stages this is the percentage of parallel components that had not responded. In order to apportion utility loss as a resource, the authors use a budget constraint of 0.1\% average utility loss across the cluster.  Our work differs from their solution in that we focus on speeding up the mature execution with which to produce answer quality. Additionally, our framework provides for provisioning of other resources based on answer quality.

Work has also been done to reduce energy cost using approximate computing~\cite{ceze,BaekCh10}.
At the application and compiler level, EnerJ~\cite{ceze} extends Java to include support for typed, approximate computing.  Baek and Chilimbi created Green~\cite{BaekCh10} specifically to replace loops and functions with approximate versions, but do so cautiously—they require the programmer to supply the approximate versions of the loops and functions, as well as the quality-of-service where they would prefer to run their application.
Baek and Chilimbi expose an interface to the programmer to replace loops and functions with approximations of the same~\cite{BaekCh10}.  Green offers statistical guarantees that the Quality of Service desired by the programmer will be met. However, it is left up to the programmer to provide approximate functions as replacements.  In the case of loops, it is left up to the programmer to provide a quality-of-service function for early termination. Green uses the output of a function compared to the output of the corresponding non-approximate function to determine the loss in quality-of-service, unless told otherwise. Using the programmer-specified inputs and functions, Green does a first-pass run of the application to determine expected loss in quality-of-service.  Before processing the full application, Green uses the information gathered from this first pass to generate a version which only uses approximation when doing so is not expected to violate the user-specified quality-of-service request~\cite{BaekCh10}.
The authors of EnerJ~\cite{ceze} combine approximate storage, approximate algorithms, and approximate computation into an easy to program interface, which separates data that requires precision from data that is allowed to be approximate. EnerJ is an extension to Java that uses approximate computing to reduce the energy consumed by a user program. The mechanism they use to keep precise data distinct from approximate data is a system of type qualifiers that allows the programmer to specify portions of their code which can be approximated without damaging quality of service. Data can be identified as permissively approximate by using a specific type identifier. It is then impossible to assign approximate values to the default precisely typed variables without endorsing these assignments explicitly. Operations are made approximate by overloading operators, and specifying that only approximately typed variables may use said approximate operators. As with overloaded operators, programmers can also specify two different algorithms for any function, one precise and one approximate, and use typing to force use of the approximate algorithm when the function results are assigned to an approximate variable.
EnerJ seeks to directly reduce energy cost using approximate computing, but neither directly quantifies the effect of its approximate computing mechanisms on overall application answer quality. EnerJ computes application-specific quality-of-service error offline for its results section~\cite{ceze}. Ubora can be used to reduce energy use of an application while keeping the answer quality at a target level. This is possible because the online answer quality trace can be used to trigger modifications to system variables such as timeouts, load shedding, reduction in auto-scaling, and DVFS scaling, all based on preset target answer quality levels.

Approximate computing is also used to tailor content for specific clients~\cite{kamat2014distributed,foxgr99,chen03,kephart-icac-2015}.
DICE~\cite{kamat2014distributed} focuses on the challenges of distributed system building for exploring data cubes interactively.  Ad-hoc analytics have been growing in popularity with end users, but the delays involved are heightened when directed at data represented as a CUBE. The framework the authors built uses speculative execution, data subsets, and faceted data cube exploration. The authors aim to deliver sub-second latencies on queries which access distributed data from billion-tuple cubes, without keeping the entire data cube in fast cached memory or using offline data sampling, which does not update in real time. This work takes advantage of the observation that queries seldom happen in a vacuum, and tend to instead be one of several related queries. In between executing queries, the authors use wait time to speculatively execute the queries most likely to be asked next, and cache these results. This work also implements timeouts on total query execution, so that even if only one of the data shards is assembled in post processing, some answer will be available. Thirdly, the work in this paper used sampling, increasing the data sampling rates for the most likely speculative queries found~\cite{kamat2014distributed}.  This paper is very similar in two ways to Ubora, in that we also uses cached data from queries hidden from the user.  CUBE uses this cached data to improve the latencies of further queries within a user session.  In our work, we have explored the possibility of improving user latency using cached results of mature executions, but our primary focus has been on using this data to measure answer quality.  CUBE is optimized to use sampled data to reduce the amount of resources spent running speculative queries.  In contrast, Ubora uses complete data for the queries it hides from the user, but only performs such mature executions for sampled queries.
Fox et al. show how a proxy-based approach could dynamically scale quality of web results across different end platforms~\cite{foxgr99}. Their work uses lossy compression to distil specifically typed information down to the portions with semantic value. Their proxy adapts on demand to fit the needs of a client.  A high end desktop machine can support downloading a larger number of bytes over the network than a smart phone; reducing the number of bytes necessary to download in order to view the page benefits such mobile devices. In contrast to our work, Fox et al. accesses a mature execution from a web server and approximates this data to meet the needs of a range of client platforms. We instead focus on services that provide online results, and measure the amount of approximation present.
Chen et al. use a thumbnail sized image of each web page as a table of contents for simplified browsing~\cite{chen03}. For pages in which splitting content into blocks is not feasible, the authors use the table of contents image to enable automatic positioning of the full-size web page on the mobile screen. Their methods introduce error in 10\% of tested web pages in order to simplify the viewing of all on mobile screens.
As in our work, Kephart and Lenchner focus on online data-intensive computations occurring across multiple components working together~\cite{kephart-icac-2015}. While we measure the answer quality of online system responses, their work displays interpreted system responses to the user and corrects computations when the user indicates incorrect analysis.

Our work focuses on data-intensive applications, and uses application-specific similarity metrics to study answer quality.  SocialTrove also focuses on data-intensive applications. Instead of measuring answer quality, SocialTrove uses application-specific similarity metrics to automatically cluster and summarize social media data~\cite{amin-icac-2015}.

Other works have focused on scheduling to increase answer quality and throughput~\cite{he-icac-2012,he-socc-2012,ren-icac-2013,zheng-cloud-2015}.
He et al. use a budget consisting of total execution time for current queries to determine whether and how long to schedule a query~\cite{he-icac-2012}. Their work uses a feedback mechanism to help ensure the desired response times are being met, and an optimization procedure to schedule based on request service demands and response quality profiles. Their algorithm takes advantage of prior knowledge regarding the overall concave quality profile of Microsoft Bing to estimate the individual request quality profile, rather than attempting to measure request quality with a mature execution.
Zeta~\cite{he-socc-2012} was designed to better schedule requests in online servers for high response quality and low response quality variance~\cite{he-socc-2012}. Zeta focuses on online services that produce partial results under a deadline, where additional computation time produces diminishing returns in additional response quality. Their response quality, like our answer quality, uses an application-specific metric to compare a partially executed request to a full execution.  They measure their response quality offline.
Ren et al. explored how heterogeneous processors can execute long requests on faster cores and shorter requests on slow cores to achieve high throughput and high quality~\cite{ren-icac-2013}. The authors implemented a new algorithm, FOF, which focuses on requests where service demands are unknown.  These new requests are scheduled on the fastest idle core.  Then, requests are migrated from slow cores to faster cores as necessary. This algorithm can improve answer quality and throughput in heterogeneous processors as compared to homogeneous processors with the same power budget. The authors used Bing without deadlines in a controlled setting to produce mature executions, and then used this data in their simulation study. We also improve metrics using adaptive resource management based on answer quality, but Ubora produces mature executions online.
Zheng et al. wrote a deadline-agnostic scheduler ISPEED for anytime algorithm workloads~\cite{zheng-cloud-2015}. Their scheduler focuses on maximizing total utility over all jobs in the cluster, and ignores concerns regarding individual query deadlines.  In addition to the utility functions used in ~\cite{he-icac-2012,he-socc-2012}, Zheng et al. also performed a user study for the Google search engine to find its average utility function~\cite{zheng-cloud-2015}.

Also highly related to Ubora is the area of adaptive resource allocation~\cite{spinner2014icpe,gandhi2014adaptive,lama2013aroma}.
Spinner et al. present their library for estimating resource demands with seven different approaches~\cite{spinner2014icpe}. Using their library, it is possible to control how often the resource use is sampled, when and for how long to perform the estimate.  In general terms, Ubora also supports estimation of resource demand for mature executions.
Gandhi et al. show that their new autoscaling cloud service, DC2, can learn an application's system parameters and scale based on its understanding of resource requirements~\cite{gandhi2014adaptive}.  The authors' solution attempts autoscaling applications without direct knowledge of their needs, instead relying on user-specified SLA information, virtual cpu statistics, and knowledge of request URLs. Like Ubora, DC2 is mostly transparent, with key information provided by the user. However, DC2 focuses directly on autoscaling rather than providing a solution for processing mature executions online.
Lama and Zhou describe their implementation of an automated resource provisioning system~\cite{lama2013aroma}.  Their system, AROMA, targets quality of service while minimizing cost, using allocation of resources in a heterogeneous cloud and Hadoop parameter configurations. However, instead of directly profiling each workload and regulating resources based on answer quality, AROMA profiles each workload for a short time on a staging cluster before matching the workload's signature to a cluster of workloads with a set of associated resources.

\section{Conclusion}
\label{sect:conclusion}

OLDI queries have complex and data-parallel execution paths that must produce
results quickly.  Data used by each query is skewed across data
partitions, causing some queries to time out and return premature results.
Answer quality is a metric that assesses the impact of
timeouts on the quality of results.  It is challenging to
compute online because it require results from mature
executions that are unaffected by timeouts.  
This paper describes Ubora, an approach to speed up mature
executions by reusing intermediate computations from online
queries, i.e., memoization.  Ubora adopts a challenging
systems-level approach that allows us to measure answer
quality for a wide range of services.  Our implementation
includes novel context tracking for commodity operating
systems and bandwidth optimizations.  The evaluation shows
that Ubora produces mature results faster than competing
transparent approaches and nearly as fast as
less flexible, application-specific approach.  Most
importantly, Ubora produces answer quality quickly enough to
enhance online system management. 

\section{Addendum}
{\small
{\noindent \bf Availability:} An extended version of this
paper, Ubora's source code and executable images of our OLDI
workloads are available at: {\em \url{http://www.cse.ohio-state.edu/~kelleyj/ubora.html}}.

{\noindent \bf Acknowledgements:} This work was supported by
NSF grants CAREER CNS-1350941 and CNS-1320071, and also by
the Oak Ridge Leadership Computing Facility managed by UT
Battelle, LLC for the U.S. DOE (contract
No. DE-AC05-00OR22725).
}

\bibliographystyle{abbrv}
{

\bibliography{bibliography}
}

\end{document}